\documentclass[5p,times]{elsarticle}
\usepackage{color}
\usepackage{lineno,hyperref}
\modulolinenumbers[5]
\begin{document}

\title{Influence of Interface Geometry on Phase Stability and Bandgap Engineering in Boron Nitride substituted Graphene: A 
Combined First-principles and Monte Carlo Study}

\author{Ransell D'Souza}%
\ead{ransell.d@gmail.com; ransell.dsouza@bose.res.in}

\author{Sugata Mukherjee}%
\ead{sugata@bose.res.in; sugatamukh@gmail.com}

\author{Tanusri Saha-Dasgupta}%
\ead{t.sahadasgupta@gmail.com}

\address{Department of Condensed Matter Physics and Materials Science \\
S.N. Bose National Centre for Basic Sciences, Block JD, Sector III, Salt Lake, Kolkata 700098, India}

\begin{abstract}
Using combination of Density Functional Theory and Monte Carlo simulation, we study the phase stability and electronic properties of two dimensional hexagonal composites of boron nitride and graphene, with a goal to uncover the role of the interface geometry formed between the two. Our study highlights that preferential creation of extended armchair interfaces may facilitate formation of solid solution of boron nitride and graphene within a certain temperature range. We further find that for band-gap engineering, armchair interfaces or patchy interfaces with mixed geometry are most suitable. Extending the study to nanoribbon geometry shows that reduction of dimensionality makes the tendency to phase segregation of the two phases even stronger. Our thorough study should form an useful database in designing boron nitride-graphene composites with desired properties.  
\end{abstract}


\maketitle

\section{Introduction}
\begin{figure*}[t]
\centering \includegraphics[width=12.2cm,keepaspectratio]{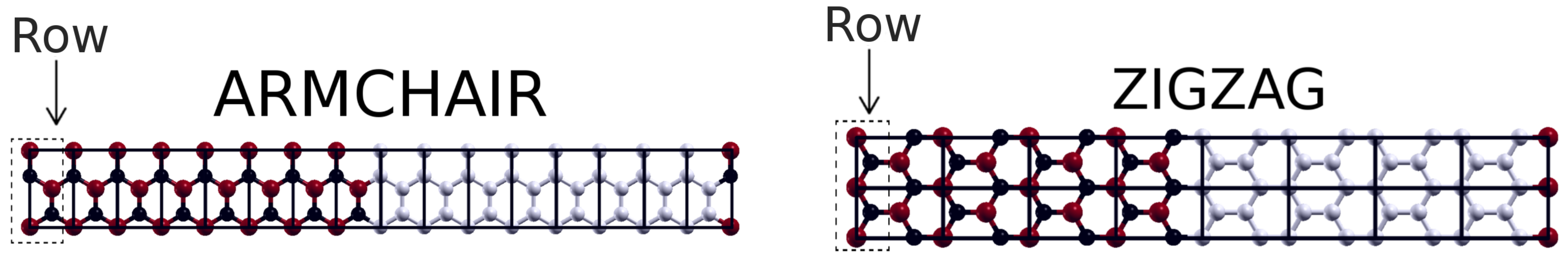}
\caption{(Color online) The arm-chair (left panel) and zig-zag interface created between h-BN and graphene within the supercell. The $16 \times 1$ armchair and $8 \times 2$ zigzag supercells generated by the orthorhombic unit cells are shown by the solid black lines. The row used for replacing C atoms
by BN atoms in two cases have been marked by the black dotted boxes. The carbon, B and N atoms are shown as white (light gray), red (dark gray) and black colored balls.}
\end{figure*}
Graphene \cite{geim04, geim07} and its structural analog, a single sheet of hexagonal boron nitride (h-BN) comprising of 
alternating boron and nitrogen atoms\cite{hBN06, hBN07} in hexagonal ring, provide prototype models for the study of two-dimensional (2D) 
systems. Besides being interesting from fundamental physics point of view, they offer technological importance for possible applications 
in the field of nanoelectronics \cite{cnrrao09, ajayan10, dean10, levendorf12}. In spite of sharing the same structural motif with only about 2$\%$
difference in their lattice constants, h-BN is an insulator with a large band-gap of more than 5 eV, while graphene is a zero-gap semi-metal
with Dirac cone band structure. It was thus thought that band gap engineering may be achieved by mixing graphene with h-BN (termed as h-CBN hereafter). 
Films of h-CBN were initially synthesized by Panchakarla et al \cite{cnrrao09} and by Ci et al \cite{ajayan10} using Chemical Vapor Deposition 
(CVD) technique, in which concentrations of C and BN could be carefully controlled. Liu et al  \cite{ajayan13} showed that planar graphene/h-BN hybrid 
can be seamlessly stitched together by growing graphene in lithographically patterned h-BN atomic layers. However, the formation of solid 
solution of graphene and h-BN is found to be thermodynamically limited, as graphene and h-BN have been reported to phase segregate both experimentally and theoretically 
\cite{ajayan10,yuge,waghmare2015,Lu,seol2011}.
The h-CBN system thus consists of segregated graphene or BN nanophases embedded in the matrix
of the other. In presence of metal support laterally joined structure of h-BN and graphene has been achieved \cite{Lu}. 

The interface formed between graphene and h-BN can be of zig-zag type or arm-chair type in case of laterally joined strip structures, or can be
of mixed type as would be the case for isolated patches. With the advancement of synthesis technique, specially on metal support, it may be
possible to synthesis samples with preferential control of one geometry of interface over another. Though there exists certain theoretical studies 
in this respect,\cite{rama2011,jungthawan2011,martins2012,cahangirov2011,berseneva2013,martins2013,kaplan2014,peng12,guilhon17} a systematic study of the influence 
of the interface geometry on the phase stability of h-CBN will be highly desirable. In the present study, we address this issue
by considering periodic array of h-BN and graphene strips, with zig-zag and armchair interfaces, formed by replacing graphene rows with boron-nitride 
rows within a given supercell. We study the phase stability of the constructed structures within the first-principles density functional theory (DFT) 
approach together with regular solid solution model.

We further investigate the microstructures formed by considering Monte Carlo (MC) simulations 
based on an underlying bond Hamiltonian with DFT derived bond energies. MC simulations also enable us to compute the spinodal line for the
MC generated interfaces with patchy structures. We additionally explore the effect of dimensionality reduction on phase stability of h-CBN. 
The miniaturization required for device applications use the so-called nanoribbon geometry with finite width in one direction and infinite in other direction.
Both graphene nanoribbons (GNR) and boron nitride nanoribbons (BNNR) have been synthesized.
Just as surface effects become predominant in 3D physics, edge effect would play a crucial role in GNR and BNNR. For example BNNR has been predicted to have narrow 
band gap and improved conductivity tuned by a transverse electric field or edge structure \cite{Chen10,zhang08}. Within the framework of MC simulation with DFT 
derived model Hamiltonian, we thus also study the phase stability properties of h-CBN in nanoribbon geometry.

Finally we study the influence of interface geometry, which can be of arm-chair type or zig-zag type formed by connection of h-BN and graphene strips,
or mixed type formed by patches of h-BN domain in the matrix of graphene or visa-versa, on the electronic structure and band gap of h-CBN system. Our 
extensive study should provide useful information on the influence of interface geometry on the phase stability and band gap engineering. 

\section{Results and Discussions}

\subsection{First-principles study of zig-zag and arm-chair interfaces between strips of h-BN and graphene}

Ab-initio DFT calculations were carried out on a  $16 \times 1$ and $8 \times 2$ orthorhombic supercell (indicated by the black solid lines in Fig.1) for the arm-chair and zigzag case respectively using the plane wave based 
Quantum Espresso code \cite{giannozzi09} \footnote{We have checked the validity of the code by comparing the calculations of structural optimization, total energies, density of states and bandstructures of pristine graphene and BN sheets with VASP \cite{vasp1} and find both codes show extremely good agreement with each other, as expected, since they provide similar computational platforms for planewave pseudopotential calculations.}. 
In these calculations, strips of h-BN and graphene 
connected either by zig-zag interface or arm-chair interface was created by replacing rows (indicated by the black dotted box in Fig.1) of C hexagons by BN hexagons of 
varying width within the unit cell, as shown in Fig 1. 
Ultrasoft pseudopotential \cite{vanderbilt90} was used to describe the core electrons and the generalized gradient approximation 
(GGA) for the exchange-correlation kernel\cite{pbe96}. A 550 eV kinetic energy cutoff for the plane-wave basis set and 2200 eV 
for the charge density was used, obtaining an accuracy of 10$^{-10}$ eV in the self-consistent 
calculation of the total energy, using a converged Monkhorst-Pack k-point grids\cite{mp76} of $6 \times 6 \times 1$.
The convergence of the computed ground state properties in terms of kinetic energy cut-off for the basis set and charge density has been checked. The positions of the atoms in the unit cell were relaxed toward equilibrium untill the Hellmann Feynman forces became less than 0.001 eV/\AA.

In order to study the phase stability of h-CBN, we first computed the cohesive energy($\Delta E$) of $({\rm C}_{2})_{x}{\rm (BN)}_{1-x}$,
which is also termed as mixing energy since it is related to the energies of the alloy related to the energies of pristine graphene and boron nitride. 
The negative value of $\Delta E$ indicates tendency to form homogeneous solid solution while positive value of $\Delta E$ indicates the tendency
to phase separate. For each concentration $x$, we calculated the mixing energy per formula unit (f.u.) of the system using DFT, which is given
by the following formula,
\begin{eqnarray}\label{coene}
\Delta  E_B &=& E\{({\rm C}_{2})_{x}{\rm (BN)}_{1-x}, a(x) \} \nonumber \\
& & -  [ x\- E({\rm C}, a_{\rm C}) + \ (1-x) \- E(h{\rm BN}, a_{h{\rm BN}})],  
\end{eqnarray}
where $E\{({\rm C}_{2})_{x}{\rm (BN)}_{1-x}, a(x) \}$ is the total energy per formula unit of $({\rm C}_{2})_{x}{\rm (BN)}_{1-x}$ at the equilibrium in-plane lattice 
constant $a(x)$; $E({\rm C}, a_{\rm C})$ and $E({\rm{\it h}BN}, a_{h{\rm BN}})$ are the total
energies per formula unit of pristine graphene and {\it h}-BN at the equilibrium in-plane lattice constants $a_{\rm C}$ and $a_{h{\rm BN}}$, respectively.
$\Delta E_B$ for different values of concentration, $x$ is shown in left panel of Fig. 2, for the arm-chair and zig-zag interface. First of all, we
find that mixing energy is positive in all cases, suggesting phase segregation between h-BN and graphene, in conformity with the literature \cite{waghmare2015, yuge,RDSM}.
Very interestingly we find that the mixing energy is substantially reduced in case of arm-chair interface compared to zig-zag interface.
This reduction
is most effective at $x$ = 0.5, for which the reduction is about 30\%.
The difference in the mixing energy between the armchair and zizag interfaces arise because of unequal number of CN and CB bonds per unit length along the interface. We have estimated the number of such bonds to be $\frac{1}{a_0}$ for the zigzag interface while for armchair interface is $\frac{2}{\sqrt{3}} \cdot \frac{1}{a_0}$, where $a_0$ is the relaxed lattice constant of $({\rm C}_{2})_{x}{\rm (BN)}_{1-x}$.

\begin{figure*}[t]
\centering \includegraphics[width=12cm,keepaspectratio]{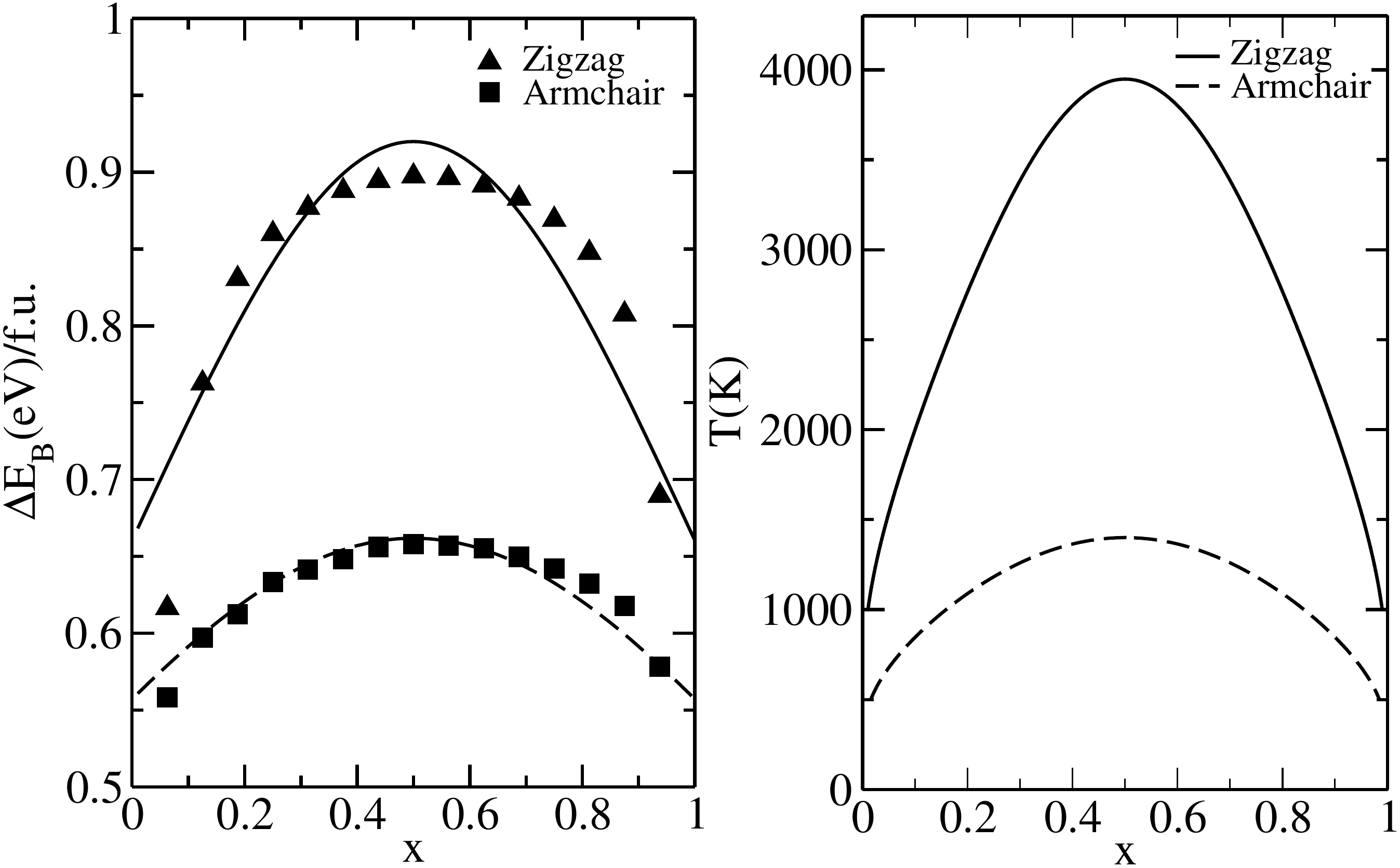}
\caption{Left panel: The mixing energy of $({\rm C}_{2})_{x}{\rm (BN)}_{1-x}$ hybrid for the arm-chair and zig-zag interfaces, plotted as a function of the
concentration, $x$. The lines are the fit of the calculated data points of the analytical form (see text). Right panel: Mean-field phase
diagram of $({\rm C}_{2})_{x}{\rm (BN)}_{1-x}$ as a function of the composition range. For each composition, the phase below the line is the segregated phase, 
while the phase above is the solid solution phase.}
\end{figure*}

From the knowledge of mixing energy, the phase stability of $({\rm C}_{2})_{x}{\rm (BN)}_{1-x}$ can be computed from a mean field approach, 
using the so-called regular solution model. The configuration entropy of mixing is defined as $S =-k_B \sum x{\rm ln}x$, where the sum runs over 
all configurations. Hence for $({\rm C}_{2})_{x}{\rm (BN)}_{1-x}$ alloys, the entropy of mixing is given by 
$S = -2k_B[x\,\ln x + (1-x)\, \ln(1-x)]$\cite{waghmare2015}, where $k_B$ is the Boltzmann constant and $x$ is the concentration of carbon. The factor 2 
arises because of the mixed occupancy of the two sublattices. The free energy is then given by  $F(T,x) = \Delta E(x) - T\,S$, where $\Delta E(x)$ is the 
mixing energy, as plotted in left panel of Fig 2. The critical temperature within the regular solution model can be obtained from the 
condition $\frac{d^2F}{dx^2} = 0$ at $x$ when $\frac{dF}{dx} = 0$. 
Fitting the mixing energy to the analytical form, $\Delta E = \frac{1}{b}\textrm{Sech}[a(x-\frac{1}{2})]$, 
it can be shown that the critical temperature will be given by
$T_C=\frac{a^2}{8bk_B}$. Fitting parameters for $x$=0.5, for arm-chair and zig-zag interfaces were found to be $a=1.208$, $b=1.511$ and $a=1.720$, $b=1.087$ 
respectively, resulting in a critical temperature of 1400 K and 3948 K. Our computed value of critical temperature for zig-zag interface is in good agreement 
with the value obtained previously in literature using cluster expansion technique and Monte Carlo,\cite{yuge} which did not take into account the specficity 
of the interface geometry. The plot of critical temperatures for the arm-chair and zig-zag interfaces for different
values of concentration $x$ is shown in the right panel of Fig 2. It follows the same trend as the mixing energy. Notably about a 65$\%$ suppression of the
the critical temperature for segregation is obtained in the arm-chair geometry of the interface at $x$ = 0.5, compared to that of the zig-zag geometry. We note
that the computed temperatures for arm-chair interfaces are substantially smaller compared to melting point of $({\rm C}_{2})_{x}{\rm (BN)}_{1-x}$
hybrids, which can be approximately estimated from the melting points of h-BN and graphene, which are about 3300 K for h-BN and 4200 K for graphene. Thus if the  $({\rm C}_{2})_{x}{\rm (BN)}_{1-x}$
hybrids can be prepared with selectively chosen arm-chair interfaces, it may be possible to arrive at a homogeneous solution of h-BN and graphene, for an appreciable range of temperature.

\subsection{Monte Carlo simulation on first-principles derived model Hamiltonian} 

{\it Setting up of the Model Hamiltonian-} DFT calculations involve large computer time even for a modestly small number of atoms and can be almost impossible for calculations involving a large number 
of atoms.  MC simulations would be ideal to deal with such situation. It is also a convenient method to know what kind of interfaces are formed if the system
is allowed to evolve without any constraint. We therefore employed Monte Carlo simulations to study the segregation of BN domains on graphene and calculate 
it's solid solution phase from the spinodal line. The Monte Carlo Simulations, within the framework of  Metropolis \cite{metropolis} algorithm, are 
based on the following Hamiltonian, defined on a bond basis with bond energies extracted out from DFT calculations.
\begin{eqnarray}
H=\frac{1}{2}\sum\limits_{i=1}^N \sum\limits_{j=1}^3 E_b(\alpha_i,\beta_j)
\end{eqnarray}
where, $N$ is the total number of atoms in the simulation cell, $E_b(\alpha_i,\beta_j)$ is the bond energy between $\alpha_i$ and $\beta_j$. $\alpha_i$ is the 
atom at position $i$, $\beta_j$ is the nearest neighbor atom. $\alpha$ and $\beta$ can be either C, B or N. The factor of 2 in the 
denominator accounts for double counting. 

In order to estimate $E_b(\alpha_i,\beta_j)$ for different kinds of bonds, which can be CC, BN, CB or CN, we first considered an isolated pair of CC, BN, CB or CN
atoms in which the C atoms are passivated with hydrogen atoms. We calculated the energy of this H-bonded pair of atoms, which is denoted as $E_{{sp}_3}$.
We then calculated the energy of the hexagonal infinite sheets built up by the same pair of atoms, which would a graphene or BN sheet if the pair of atoms is CC
and BN. For rest of the combinations, these are artificial computer-generated sheets. Energy of this infinite sheet is denoted as $E_{inf}$. Considering the
case of graphene, and considering the fact that energy of the infinite sheet in the unit cell is given by the energy of two isolated C atoms ($C_{iso}$) and the 
bond energy of C-C, $E_b(CC)$, we have,
\begin{eqnarray}
E_{inf} = 2C_{iso}+nE_{b}({\rm CC})
\end{eqnarray}
where $n$ is the  number of bonds on each carbon atom.
Similarly the energy of hydrogen passivated C-C pair can be also expressed as a sum of energy of isolated atoms and bond energies. Thus,
\begin{eqnarray}
E_{{sp}_2} = 2C_{iso}+E_{b}({\rm CC}) + 4H_{iso} + 4E_{b}({\rm CH})
\end{eqnarray}
taking into account of the fact that there are 6 hydrogen atoms required to passivate the two carbon atoms completely.
From the above two equations, one can arrive at a definition of CC bond energy as,
\begin{eqnarray}
E_{b}({\rm CC})=\frac{E_{inf}-(E_{{sp}_2}-4H_{iso} - 4E_{b}({\rm CH}))}{n-1}
\end{eqnarray}
The bond energies of other pairs can be defined similarly. Table \ref{tab}, lists the bond energies for different pairs, calculated using
DFT computed values of $E_{inf}$, $E_{{sp}_3}$ and $A_{iso}$, the latter being isolated atom energies, with $A$ = C/B/N.

\begin{table}
\caption{Calculated Bond energies of h-CBN infinite sheet (middle column) and zig-zag nanoribbon (right column) for 
different pairs (left column), obtained from DFT calculations. Similar values are obtained for arm-chair nanoribbon}\label{tab}
\begin{tabular}{ccc}
\textrm{Type} & $E_{b}(eV)$  & $E_{b}(eV)$  \\
\ & Bulk & {\scriptsize Outer most row of nanoribbon} \\
\hline
CC & -0.919 & -1.30\\
BN & -0.921 & -1.45\\
CB & -0.654 & -1.27\\
CN & -0.314 & -0.85\\
\end{tabular}
\end{table}

As is evident from the table, the bond energies for CC and BN bonds are far more stronger compared to CB and CN bonds. The random mixing
of graphene and h-BN would require formation of CB and CN bonds, which clearly is unfavorable, explaining the observed segregation behavior
in normal condition. 

Having computed the bond energies in a DFT derived way which are the input to the MC simulation for the infinite sheet of $({\rm C}_{2})_{x}{\rm (BN)}_{1-x}$,
we proceed to calculate the same for cases when the systems are nanoribbons. As can be anticipated, the bond energies of the pair of atoms
positioned near the edge of the ribbon will be different from that inside the ribbon. This effect may also extend to atoms adjacent to
the edge. Thus to compute the position-dependent bond energies, we proceed as follows. We passivate the edges of the nanoribbon with hydrogen 
atoms. For zig-zag (arm-chair) edged ribbon there are 2 (4) such H atoms in the unit cell. We built up the nanoribbons of increasing width by adding rows of atoms
along the lateral dimension of the ribbon. At each stage, two rows were added which amounts to one unit cell (u.c.). Considering the graphene 
nanoribbon with smallest width which consists of two rows of atoms, the energy of the H-passivated system is given by
\begin{eqnarray}
E_{2row} = 4C_{iso}+ 7E_{b1}({\rm CC}) + 2H_{iso} + 2E_{b}({\rm CH})
\end{eqnarray}
where $E_{b1}({\rm CC})$ is the CC bond energies of the carbon atoms belonging to the smallest possible nanoribbon. From knowledge of DFT energies
for $E_{2row}$, $C_{iso}$, $H_{iso}$, $E_{b}({\rm CH})$, the bond energy $E_{b1}({\rm CC})$ is estimated. Adding two additional rows of carbon
atoms lead to the energy given by,
\begin{eqnarray}
\hspace{-2.5em} E_{4row} = 8C_{iso}+ 8E_{b1}({\rm CC}) + 7E_{b2}({\rm CC}) + 2H_{iso} + 2E_{b}({\rm CH})
\end{eqnarray}  
Inputting the estimate of $E_{b1}({\rm CC}$ obtained from previous calculation of $E_{2row}$ gives the estimate of $E_{b2}({\rm CC})$ which is the CC bond energies of the carbon atoms immediately adjacent to rows belonging to the edges.
This process is continued to extract the row-dependent bond energies in the nanoribbon geometry.

Our computed bond energies for the nanoribbons show (see Table \ref{tab}) the bond energies to be significantly larger at the edges which converge
to the corresponding bulk values beyond five rows of atoms starting from the edge, with only marginal difference between zig-zag and arm-chair cases.
\vskip .2in

{\it MC snapshots-} Considering the case of infinite $({\rm C}_{2})_{x}{\rm (BN)}_{1-x}$ sheet, the achieved equilibrium configurations at room temperature 
are shown in Fig 3, for $x$ = 0.2, 0.4, 0.6 and 0.8. In all cases, we find that depending on the concentration, the final configuration consists of either nanopatches of C atoms embedded in matrix of BN atoms or BN nanopatches, embedded in the matrix of carbon, suggesting the typical case of segregation. The edges formed between BN and C atoms turned out to be of mixed character with
dominance of zig-zag boundaries over arm-chair. Thus, unless enforced through special growth condition, the $({\rm C}_{2})_{x}{\rm (BN)}_{1-x}$ sheet tends
to segregate with creation of edges with mixed character, as observed in initial experiments.\cite{ajayan10} We further find from the snapshots that 
in case of ribbon geometry the BN atoms tend to segregate at the edges forming extending phase segregated domains running along the lateral direction of
the ribbon, with carbon atoms in general positioned towards the central part of the ribbon. Fig 4 shows the achieved equilibrium configurations for
a arm-chair and zig-zag edged nanoribbon at room temperature for $x$ = 0.2, 0.4, 0.6 and 0.8.

\begin{figure*}[!htb]
\centering \includegraphics[width=11cm,keepaspectratio]{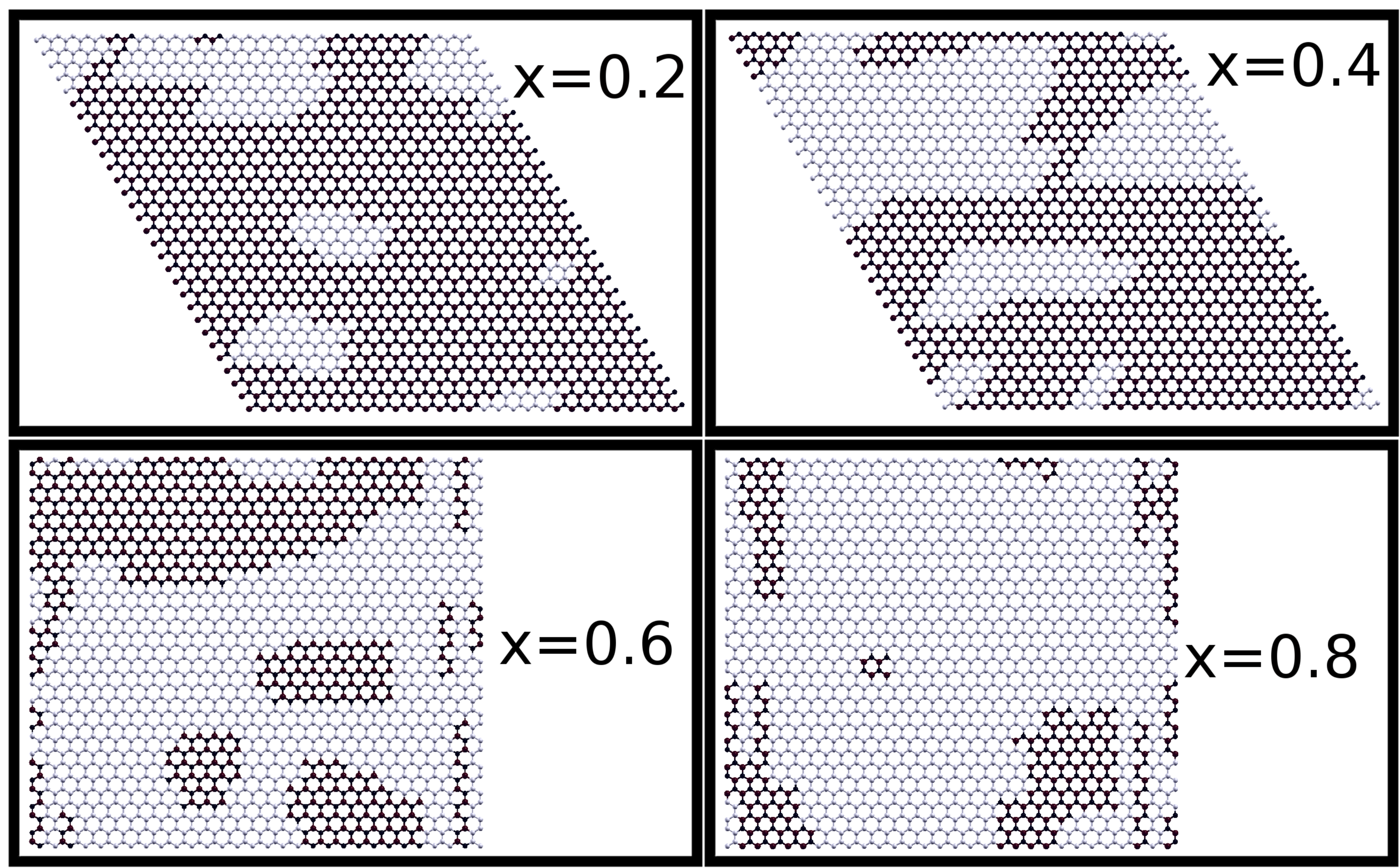}
\caption{(Color online) Snapshots of equilibrium configuration obtained in MC simulations at T = 300 K for $({\rm C}_{2})_{x}{\rm (BN)}_{1-x}$ 
infinite sheet using an hexagonal (above) and orthogonal (below) super-cell with $x$ = 0.2, 0.4, 0.6 and 0.8. The color convention of the balls is same as in Fig 1.}
\end{figure*}

\begin{figure*}[!htb]
\centering \includegraphics[width=12.5cm,keepaspectratio]{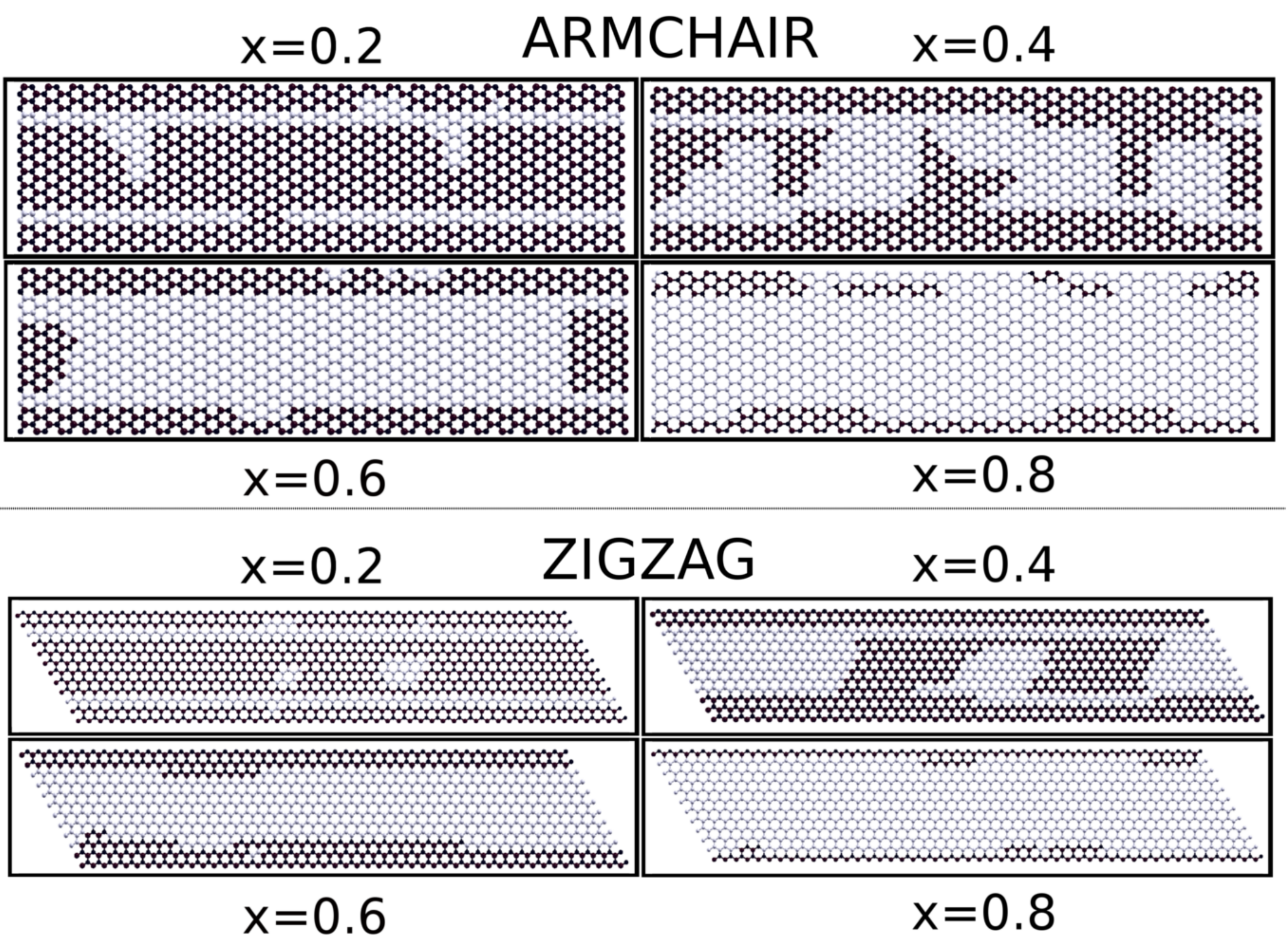}
\caption{(Color online) Snapshots of equilibrium configuration obtained in MC simulations at T = 300 K for an arm-chair and zig-zag edged $({\rm C}_{2})_{x}{\rm (BN)}_{1-x}$ 
nanoribbon with $x$ = 0.2, 0.4, 0.6 and 0.8. The color convention of the balls is same as in Fig 1.}
\end{figure*}

{\it Calculation of Spinodal line-} In order to calculate the spinodal line based on MC results, we first defined a suitable order parameter in the 
following manner. If the number of nitrogen nearest neighbors of boron is $n_b$, we define an order parameter on each boron to be $\eta_{p_b} = \frac{n_b}{3}$. 
Similarly, if the number of boron nearest neighbors of nitrogen is $b_n$ , we define an order parameter on each nitrogen to be $\eta_{p_n} = \frac{b_n}{3}$. 
We then define an average order parameter $\langle \eta \rangle$ for the system as $ \langle \eta \rangle = \frac{1}{N_B + N_N} \sum_{i \in B,N} \frac{\eta_{p_b}+\eta_{p_n}}{2}$, which is averaged over all the Nitrogen ($N_N$) and Boron ($N_B$) atoms
in the simulation cell. We found that  at low temperatures the value of the order parameter increases as the system evolves from an initial random 
configuration to a final configuration, while at higher temperatures the order parameter evolves close to that of the initial random configuration.
After a critical temperature the system accepts any exchange of BN and C dimers keeping the order parameter the same as the random configuration. We defined 
this temperature as the critical temperature. We repeated this procedure for various concentrations thus obtaining the spinodal line. The points above the
spinodal line refers to the disordered solid solution phase while those below refers to the segregated phase. The left panel of Fig 5 shows the
spinodal line for the infinite sheet of $({\rm C}_{2})_{x}{\rm (BN)}_{1-x}$, and the nanoribbons of $({\rm C}_{2})_{x}{\rm (BN)}_{1-x}$ having width of 8 u.c. with zig-zag and arm-chair edges. 
For the infinite sheet, results were obtained for about 20000 atoms in the periodic unit cell, and 2 $\times$ 10$^{5}$ MC steps were used to reach the equilibrium.
For the nanoribbons, the number of atoms in the lateral direction with periodic boundary condition was chosen to be 10000. 
We find the critical temperature of phase segregation is substantially high in case of ribbons compared to the infinite sheet, which makes it
comparable to the corresponding melting point. 
This observation that the critical temperature of ribbons being larger than that of infinite sheets is rationalized by the strengthening of bond strengths at edges compared to bulk values (See Table \ref{tab}).  
In case of infinite sheet, with interfaces of mixed character the calculated transition temperatures
though less than that of the ribbon geometries are high enough, prohibiting mixed solution of h-BN and graphene under normal condition, unless
the concentration is very low. In the right panel, we show the variation of the critical temperature at $x$ = 0.5,
as a function of the width of the nanoribbon for both zig-zag and arm-chair cases. We find the transition temperature for arm-chair is systematically less
than that of the zig-zag nanoribbon with the difference being larger for ribbons of smaller widths. This is in line with our observation from mean-field study 
of infinite sheet that transition temperatures are suppressed in case of arm-chair interface compared to that of zig-zag interface.

\begin{figure*}[!htb]
\centering \includegraphics[width=12cm,keepaspectratio]{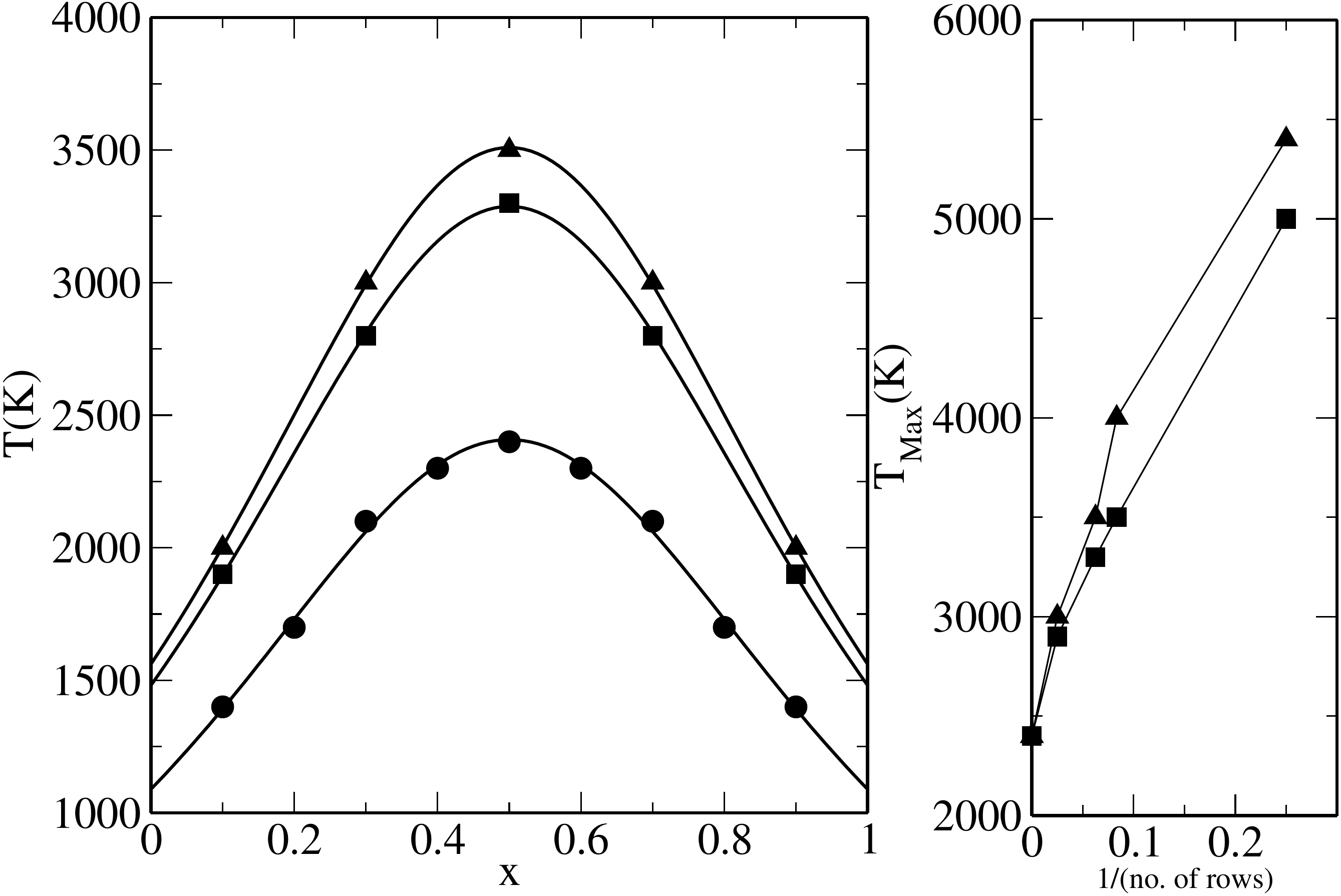}
\caption{Left panel: Spinodal lines for the infinite (circles) sheet, arm-chair (square) as well as zig-zag (triangle) edged nanoribbon of $({\rm C}_{2})_{x}{\rm (BN)}_{1-x}$, calculated by MC simulation. Shown are the data for nanoribbons of width 8 u.c. The lines are guide to eye. Right panel: Calculated transition temperature at $x$ = 0.5 plotted as a function of the inverse of the width of the nanoribbon, for the zig-zag (triangle) and arm-chair (square) edged nanoribbons. The width is measured in terms of number of row of atoms counted along the transverse dimension of the ribbon.  Two rows of atoms constitute a unit cell. The data point at zero of the x-axis corresponds to the value obtained for the infinite sheet.}
\end{figure*}

\subsection{Bandgap Engineering in $({\rm C}_{2})_{x}{\rm (BN)}_{1-x}$}

\begin{figure*}[t]\label{bs}
\centering \includegraphics[width=13cm,keepaspectratio]{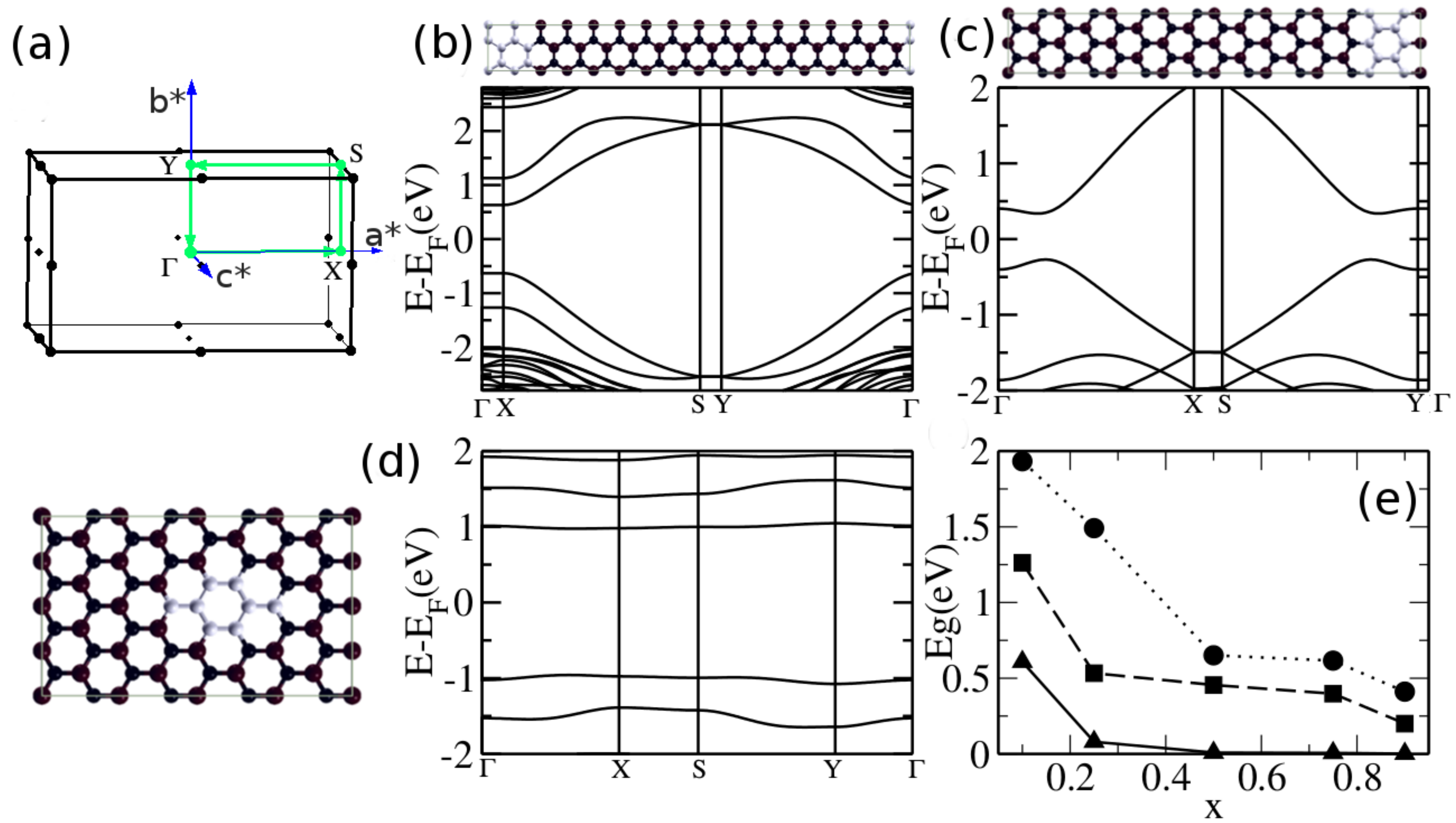}
\caption{(Color online) (a) Orthorhombic BZ with high symmetry points. (b,c,d): Band structure of $({\rm C}_{2})_{x}{\rm (BN)}_{1-x}$ with $x$ = 0.1, plotted along the high-symmetry points of the BZ corresponding to the orthorhombic cell.  (b,c) show the band structure for arm-chair and zig-zag 
interfaces, while (d) shows the band structure for the mixed interface. The corresponding interface geometries are shown by the side of band structure plots. The color convention of the balls in these figures is same as in Fig 1. (e): The band gap values plotted as a function of varying concentration, $x$. The circles, squares and triangles represent the data corresponding to mixed, arm-chair and zig-zag interfaces.}
\end{figure*}

Finally, we investigate the influence of the interface geometry on the band gap engineering. In the study described so far, we have considered
selectively created zig-zag type interface, or arm-chair type interface, and freely evolved interface generated in MC simulation which turned
to be of mixed kind. Therefore a systematic study of electronic structure of the composite structures possessing these different interfaces i.e. armchair, zigzag and patches (mixed combination of zigzag 
and armchair) is important and necessary. These results may throw light on the underlying physics of band-gap engineering in optoelectronic devices.

In the upper middle, upper right and lower middle panels of Fig 6, we show the band structure of $({\rm C}_{2})_{x}{\rm (BN)}_{1-x}$ hybrids considering the zig-zag, arm-chair and patchy interfaces for $x$ = 0.1. In lower
right most panel we also show the variation of band gap as a function of varying concentration $x$ for each of these interface geometries. 
From the three band structure plots it is evident that in all different cases of interface geometries the band-gaps are direct band-gaps and hence an electron 
can directly emit a photon without a change in momentum, giving such materials a high optical absorption. This aspect continues for other $x$ values as well.
A significant difference in the bandstructure of patchy interface compared to that of zigzag or armchair interfaces is that the bands are almost flat 
in case of patchy structure, while there is appreciable dispersion for the zig-zag or arm-chair interface, arising due to extended connectivity. This in turn implies the quenching of kinetic energy 
of electrons for the patchy interfaces amounting localization of the electrons at the states close to valence band maximum (VBM) and conduction band minimum (CBM). 
For the plot of the band gap, we further find that for a given concentration the band-gaps depend crucially on for the interface geometry. 
Since the band along YS (XS) direction in the arm-chair (zig-zag) interface is flat, we believe that the graphene nanoribbon geometry embedded in the CBN sheet mainly determines the entire low-energy band structure \cite{son06}.
For a zig-zag interface, the closing of the band gap and a metallic behavior is obtained already for a concentration of $x$ = 0.5, with significant suppression of band-gap
at a concentration of $x$ = 0.25. On the other hand, both for arm-chair and mixed interfaces, band gap is not closed even with large substitution of carbon
atoms e.g. $x$ = 0.9, giving rise to a large concentration window available for band gap tuning.
 
\section{Summary and Discussion}
To summarize, we have studied theoretically the influence of various geometrical shapes of the interfaces formed between phase segregated graphene and 
h-BN on the properties of h-BN substituted graphene systems. We have employed for this purpose mean-field regular solution model as well as Monte
Carlo simulations of first-principles derived models. Our calculations show a rather strong dependence of the interface geometry both on the phase
stability and the band gap engineering, the latter being the original motivation for studying graphene-BN hybrid systems. 

We found a significant suppression of the segregation temperature is obtained for the arm-chair shaped interfaces, giving rise to the possibility of 
achieving homogeneous solution of graphene-BN alloy phase if extended arm-chair interfaces can be created selectively. We further found achieving such homogeneous 
solution phase becomes progressively difficult upon reduction of dimensionality, in moving from infinite sheet to nanoribbons of $({\rm C}_{2})_{x}{\rm (BN)}_{1-x}$
of smaller and smaller widths. 

Our study on band structure showed for band gap tuning arm-chair or mixed interfaces are better candidates compared to zig-zag interface. For the later
the gap is significantly reduced and closes completely beyond a substitution limit of 0.5. On the other hand, the band gap remains finite for the arm-chair
or mixed interfaces even for high level of substitution BN by carbon atoms of 0.9 (the highest value studied in the present study). Our band gaps have been 
calculated using GGA exchange-correlation functional which is expected to underestimate the values of the band gap, but the calculated trend should be 
robust, as has been shown in the study employing both hybrid functional and GGA functional.\cite{Zhu}

Finally, our thorough and extensive study considering different possible interfaces in bulk as well as reduced dimensionality in $({\rm C}_{2})_{x}{\rm (BN)}_{1-x}$  
composite systems should provide an useful insight on the interfacial geometry effect on properties. Given the experimental possibility of control on phase stability
of $({\rm C}_{2})_{x}{\rm (BN)}_{1-x}$ composite systems using supported and patterned substrates,\cite{Lu} it might be possible to selectively create interface
of one type over other with desired properties. In this context, band structure and stability of various isomers of 2D infinite sheet of 
(BN)$_m$(C$_2$)$_n$ composites have been studied from the view point of chemical concepts of conjugation and aromaticity,\cite{Zhu} which also pointed out 
that the relative widths and arrangement of graphene phases in embedded h-BN matrix in an infinite sheet will be crucial in realizing BN substituted 
graphene systems with desired band gap. Our alternative approach of study reconfirms that idea, and additionally shows the effect of reduced dimensionality
which is detrimental to achieve homogeneously mixed state.

\section{Acknowledgments}
One of us (RD) acknowledges S.N. Bose National Centre for Basic Sciences for a Senior Research Fellowship. All calculations were performed in the High Performance
Cluster computer platform at S.N. Bose Centre. 

\vskip1cm
\bibliographystyle{elsarticle-num}

\end{document}